# Advanced One-Parameter CKM Mixing Matrix and Universal Deviation from Exact Quark-Lepton Complementarity

E. M. Lipmanov
40 Wallingford Road # 272, Brighton MA 02135, USA

**Abstract**

In this paper, realistic CPT-symmetric united at substance quark and neutrino mixing matrices are studied by the idea that they are shifted from respectively unit and bimaximal benchmark matrices by one new small universal empirical $\varepsilon$–parameter. At leading finite $\varepsilon$–approximation quark-lepton complementarity (QLC) is an exact regularity. *Equal* solar and atmospheric deviations from exact QLC constitute one of the main inferences at next to leading approximation. That universal violation of exact QLC advances the status of the QLC idea; and the more so, as that deviation from QLC is quantitatively estimated and used for accurate calculations of lepton mixing angles. Quark and neutrino Dirac CP-violating phases are determined by the considered quadratic hierarchy paradigm in flavor phenomenology. Inferences are partly supported by quark heavy flavor unitarity triangle angles and testable at B-factory and LHC b experiments. Estimated magnitude of the unitarity triangle gamma-angle concurs with the quark CP-violating phase to within $\sim 5\times 10^{-4}$. The final quark CKM mixing matrix in terms of the $\varepsilon$-parameter is in excellent quantitative agreement with world data and suggests a fitting form of neutrino PMNS mixing matrix. The $\varepsilon$-constant implications in flavor phenomenology are considered.

# 1. Introduction

Flavor physics is a fundamental part of elementary particle physics, but in contrast to the one-generation Standard Model with three particle mass copies there is yet no well established flavor theory that would enable high-accurate calculations of flavor quantities, such as particle mixing angles, despite of the large and growing number of relevant accurate experimental data and many symmetry based flavor mixing models. As goes without saying, physics is still an experimental science; empirical data may suggest substantially new physical ingredients, as e.g. a possible basic dimensionless parameter, and may point to simple accurate quantitative regularities between different flavor quantities [1] that lead to useful new flavor phenomenology which may agree or disagree with the symmetry approach.

It should be mentioned that stimulating phenomenological two-steps approach to quark and neutrino mixing matrices was developed by W. Rodejohann [2] where the unit quark matrix and bimaximal neutrino one are considered as 'reference' matrices and the realistic deviations from that reference are described by an empirical small $\lambda$–parameter approximately equal to the magnitude of the quark Cabibbo mixing angle $\theta_C$.

Encouraging approximate agreement with quark and neutrino mixing patterns is achieved [3], [4] by supplementing the extreme form of exact quark-lepton complementarity[1] (QLC) of the benchmark[2] mixing matrices with *equal,* maintaining exact QLC quark and neutrino deviations from benchmark in terms of a small empirical $\varepsilon$–parameter. This newly introduced $\varepsilon$-parameter seems pertinent for revealing accurate quantitative inferences from experimental data in particle flavor-electroweak phenomenology. It is ~ 3 times smaller than $\lambda$ and is accurately defined by its empirical relation to the fine structure constant at zero momentum transfer:

$$\varepsilon \cong 0.082085 \cong \exp(-5/2). \qquad (1)$$

---

[1] Quark-lepton complementarity is much studied earlier e.g. [5].
[2] Benchmark mixing matrices coincide with the two reference-matrices of ref.[2]. Detailed definition of quark and lepton benchmark flavor patterns is given in [3], where maximal neutrino mixing is related to the approximation of exact mass degeneracy while zero quark mixing is related to infinitely large benchmark mass hierarchies.

It was observed that at leading approximation different powers of the $\varepsilon$–parameter provide fitting estimations of charged lepton, quasi-degenerate neutrino and quark mass ratios, neutrino oscillation hierarchy parameter and quark and neutrino mixing matrices.

In this paper, advanced substantially united and complete, quark and neutrino mixing matrices are established mainly by three observed regularities 1) *Equal-form* equations for the three independent CKM [6] mixing angles in terms of $\sin^2$-double-angle presentation and the small $\varepsilon$-parameter[3], 2) *Equal* realistic solar and atmospheric deviations from exact QLC in terms of such presentation and $\varepsilon$-parameter, and 3) Quark and neutrino Dirac CP-violating phases and (1-3) mixing angles are defined by the considered earlier quadratic hierarchy paradigm in flavor phenomenology [3]-[4]. Both leading flavor regularities – QLC and quadratic hierarchy for generic flavor-mixing pairs – are exact empirical regularities only at leading $\varepsilon$-approximation; both ones are weakly violated at next to leading $\varepsilon$-approximation.

To the author's knowledge, high-accurate one-parameter presentation of the Cabibbo-Kabayashi-Maskawa mixing matrix in quark mixing phenomenology is obtained without particular adjusting parameters for the first time. As heuristic means, that accurate CKM-matrix leads to a prediction of fitting neutrino PMNS mixing matrix and eventually to the notion that the suggested by data $\varepsilon$-parameter presents a glimpse on new flavor physics ingredient.

---

[3] Different motivations that lead to definition of the $\varepsilon$–parameter are considered in [4] and [3] and references therein.

In Sec.2 accurate realistic quark mixing matrix is obtained as small $\varepsilon$-shift from minimal benchmark mixing matrix. In Sec.3 universal realistic small $\varepsilon$-deviation from exact quark-lepton complementarity is formulated. In Sec.4 neutrino mixing matrix is obtained as small $\varepsilon$-shift from bimaximal benchmark mixing matrix. The conclusions are in Sec.5.

## 2. Accurate quark mixing matrix as small shift from unit benchmark mixing one

Benchmark quark mixing pattern [3] describes minimal (zero) quark mixing at zero value of the universal parameter, $\varepsilon = 0$, in terms of $\sin^2$-double-angle presentation as given by

$$\sin^2(2\theta_c) = 0, \quad \sin^2(2\theta_{23}) = 0, \quad \sin^2(2\theta_{13}) = 0, \qquad (2)$$

where $\theta_c = \theta_{12}$ is the Cabibbo mixing angle and $\theta_{23}$, $\theta_{13}$ are the two next quark mixing ones. From the definition (2) follows the quark benchmark unit mixing matrix

$$\begin{pmatrix} 1 & 0 & 0 \\ 0 & 1 & 0 \\ 0 & 0 & 1 \end{pmatrix}_q. \qquad (3)$$

At finite but small parameter $\varepsilon$ the primary conditions (2) should be transformed to

$$\sin^2(2\theta_c) = \varepsilon f_1(\varepsilon), \quad \sin^2(2\theta_{23}) = \varepsilon f_2(\varepsilon), \quad \sin^2(2\theta_{13}) = \varepsilon f_3(\varepsilon), \qquad (4)$$

where $f_i$, $i=1,2,3$, are finite functions of $\varepsilon$.

Discrete assumptions that the three functions are of equal exponential form $\sin^2(2\theta_i) = f(x_i)$ and vanish at $\varepsilon = 0$ in accordance with the pattern $f(x_i) \to x_i$ considered earlier [4], lead to the interesting semiempirical discovery that all three relations (4) are represented by the single exponential function

$$f(x) \equiv x e^x,$$

$$\sin^2(2\theta_i) = f(x_i), \quad x_1 = 2\varepsilon, \quad x_2 = \varepsilon^2, \quad x_3 = \varepsilon^4. \tag{5}$$

Notice that the deviation of the sequence $x_i$ in (5) from geometrical form is by coefficient '2' in $x_1$ that is determined by the condition of exact quadratic hierarchy for (1-2) and (2-3) mixing angles at leading $\varepsilon$-approximation [4].

The final result for the three independent CKM mixing angles $\theta_1 = \theta_c$, $\theta_2 = \theta_{23}$, $\theta_3 = \theta_{13}$ is given by

$$\sin^2(2\theta_c) = f(2\varepsilon), \quad \sin^2(2\theta_{23}) = f(\varepsilon^2), \quad \sin^2(2\theta_{13}) = f(\varepsilon^4),$$

$$\theta_c \cong 13.047°, \quad \theta_{23} \cong 2.362°, \quad \theta_{13} \cong 0.193°. \tag{6}$$

In terms of the standard parameterization [6],

$$V \cong \begin{pmatrix} C_{12}C_{13} & S_{12}C_{13} & S_{13}e^{-i\delta} \\ -S_{12}C_{23}-C_{12}S_{23}S_{13}e^{i\delta} & C_{12}C_{23}-S_{12}S_{23}S_{13}e^{i\delta} & S_{23}C_{13} \\ S_{12}S_{23}-C_{12}C_{23}S_{13}e^{i\delta} & -C_{12}S_{23}-S_{12}C_{23}S_{13}e^{i\delta} & C_{23}C_{13} \end{pmatrix}, \tag{7}$$

with quark mixing angles from (6), accurate quantitative prediction for the quark CKM mixing matrix is obtained,

$$V_q \cong \begin{pmatrix} 0.97418 & 0.22575 & 0.00337e^{-i\delta} \\ -0.22556-0.00014e^{i\delta} & 0.97336-0.00003e^{i\delta} & 0.04121 \\ 0.00930-0.00328e^{i\delta} & -0.04015-0.00076e^{i\delta} & 0.999145 \end{pmatrix} \tag{8}$$

where $\delta = \delta_q$ is the magnitude of CP-violating K-M phase,

$$\delta_q \cong 65.53°, \tag{9}$$

as a unique solution of the quadratic hierarchy equation [4]

$$\sin^2 \delta_q = 2\cos \delta_q, \quad \cos \delta_q = (\sqrt{2}-1). \tag{9'}$$

Concise description of the quark mixing matrix (8) is analytically described in terms of the ε-parameter by the four equations (6) and (9′).

It should be noticed that necessary violation of CP-invariance in the CKM mixing matrix (8) is a result of the general quadratic hierarchy regularity for generic pairs of flavor quantities [4a].

From the result (8) follows the matrix form 'in moduli'

$$V_a \cong \begin{pmatrix} 0.97418 & 0.22575 & 0.00337 \\ 0.22556 & 0.97336 & 0.04121 \\ 0.00848 & 0.04044 & 0.999145 \end{pmatrix} \quad (10)$$

that is in very close agreement with the PDG presentation of global fit in the SM [7],

$$V_{CKM} \cong \begin{pmatrix} 0.97419 \pm 0.00022 & 0.2257 \pm 0.0010 & 0.00359 \pm 0.00016 \\ 0.2256 \pm 0.0010 & 0.97334 \pm 0.00023 & 0.0415 \pm 0.0010 \\ 0.00874 \pm 0.0003 & 0.0407 \pm 0.0010 & 0.999133 \pm 0.000044 \end{pmatrix} \quad (11)$$

Most of the matrix elements in (10) are within 1 S.D. ranges with one exception: $V_{ub} = S_{13} = 0.00337$ is about 1.4 S.D. from the PDG best-fit value 0.00359 from (11).

The definition of the CP-violating phase (9) is partly supported by the estimations of flavor quantities related to unitarity triangle angles,

$$\beta = \arg(-V_{cd}V^*_{cb}/V_{td}V^*_{tb}),$$
$$\alpha = \arg(-V_{td}V^*_{tb}/V_{ud}V^*_{ub}), \quad \gamma = \arg(-V_{ud}V^*_{ub}/V_{cd}V^*_{cb}), \quad (12)$$

that define the actual unitarity triangle angles

$$\beta \cong 20.6°, \quad \alpha \cong 93.9°, \quad \gamma \cong 65.5°, \quad \alpha + \beta + \gamma = \pi, \quad (13)$$

$$\sin 2\beta \cong 0.659. \quad (14)$$

These estimations agree with the inferences from heavy flavor BABAR and Belle experimental data analysis [7],

$(\sin 2\beta)_{exp} = 0.681 \pm 0.025$, $\alpha_{exp} = (88 \pm 6)°$, $\gamma_{exp} = (77 \pm 30)°$, (15)

to within 1 S.D.

Notice an accurate quantitative coincidence between the magnitudes of the unitarity angle $\gamma$ in (13) and CP-violating phase $\delta_q$ (9), $(\delta_q - \gamma)/\delta_q \cong 6 \times 10^{-4}$; it is an inference from the very small relative imaginary part of the matrix element $V_{cd}$ in (8). As a result, accurate determination of the $\gamma$-angle (13) at B-factories and LHCb experiments, will directly give the magnitude of the quark CP-violating phase $\delta_q$ with predicted high precision $\sim 5 \times 10^{-4}$.

The idea of small $\varepsilon$-deviations from the unit quark benchmark mixing matrix (3) with regularities (5)-(6) quantitatively deciphers the empirical massage of small deviation of the quark CKM mixing matrix from unit matrix. It may be considered as a hint on $\varepsilon$-parameter-based new pre-SM physics of quark mixing, which appears to be stable against SM radiative corrections, comp. [8].

The established quark mixing matrix (8) leads to the estimation of the Jarlskog invariant [9]

$$J \cong 3.017 \times 10^{-5},$$ (16)

that is an important observable CP-violation quantity; inference (16) is consistent with the PDG [7] result, $J = (3.05 + 0.19 - 0.20) \times 10^{-5}$.

## 3. Small universal deviations from exact quark-lepton complementarity

Exact quark-lepton complementarity for the two largest quark and neutrino mixing angles would mean conditions

$\sin^2(2\theta_c)/\cos^2(2\theta_{sol}) = 1$, $\sin^2(2\theta^q_{23})/\cos^2(2\theta_{atm}) = 1$. (17)

Taking for granted the best-fit values of neutrino mixing angles [10]-[12],

$$(\sin^2\theta_{sol})_{exp} = 0.312 + 0.063 - 0.049, \quad (\theta_{sol})_{bf} = 33.96°, \quad (18)$$

$$(\sin^2\theta_{atm})_{exp} = 0.466 + 0.178 - 0.135, \quad (\theta_{atm})_{bf} = 43.05°, \quad (19)$$

and using the values of quark mixing parameters from the matrix (8), deviation from exact quark-lepton complementarity in terms of the $\varepsilon$-parameter is envisaged in the form

$$\sin^2(2\theta_c)/\cos^2(2\theta_{sol}) \cong \sin^2(2\theta^q_{23})/\cos^2(2\theta_{atm}) \cong \exp(4\varepsilon). \quad (20)$$

In terms of this minimal suggestion the deviation of neutrino mixing from exact QLC is about 15%.

Some phenomenologically attractive features of the suggestion (20) are 1) Universal deviations from QLC, i.e. the deviations from QLC of the atmospheric and solar mixing angles are almost equal, 2) Simple exponential form of the deviation from QLC without new parameters and 3) Deviations from QLC for neutrino mixing angles are generated only at next to leading $\varepsilon$-parameter approximation.

In terms of benchmark mixing these features read: 1) At zero $\varepsilon$-approximation, the quark and neutrino mixing matrices are equal respectively to the benchmark unit and bimaximal ones with extreme form of exact QLC,

2) At leading nonzero $\varepsilon$-approximation exact QLC is maintained. At leading $\varepsilon$-approximation the $\sin^2$- and $\cos^2$-double-angle mixing quantities of quark and neutrinos obey exact quadratic hierarchy conditions, e.g.

$$\sin^2(2\theta_c) = 2\sin(2\theta^q_{23}), \quad \cos^2(2\theta_{sol}) = 2\cos(2\theta_{atm}),$$

3) Beyond the leading $\varepsilon$-approximation the quark and neutrino deviations from benchmark mixing are different, but with equal solar and atmospheric deviations from exact QLC (20).

The considered deviation from exact QLC seems *natural* since exact QLC is generated by benchmark mixing, maintained at leading approximation of the small $\varepsilon$–parameter, deviated at next to leading $\varepsilon$–approximation, and the more so, as that deviation is universal and quantitatively estimated by relations (20) in agreement with data.

As an inference, solar and atmospheric neutrino angles must be deviated from maximal mixing $\pi/4$ in agreement with (18)-(19), but small, though essential, disagreement with the tribimaximal mixing [13].

The quantitative result for solar mixing angle from (20) is given by

$$\cos^2(2\theta_{sol}) \cong (2\varepsilon)\exp(-2\varepsilon), \quad \theta_{sol} \cong 34.042°. \quad (21)$$

For the atmospheric mixing angle, from (20), the result is

$$\cos^2(2\theta_{atm}) \cong (\varepsilon^2)\exp(\varepsilon^2)\exp(-4\varepsilon),, \quad \theta_{atm} \cong 42.996°. \quad (22)$$

From (21) follows $\sin^2\theta_{sol} \cong 0.313$ and from (22) $\sin^2\theta_{atm} \cong 0.465$; the agreement with the best-fit values (18) and (19) is remarkable.

To conclude, the exponential presentation (20) for universal deviation from exact quark-lepton complementarity leads to neutrino mixing angles that are in better then per cent agreement with the best-fit values for neutrino solar and atmospheric mixing angles [12], (18)-(19), and are compatible with other analyses [14]. So, the realistic QLC condition is a result of extreme form of exact QLC of the starting quark and neutrino reference (benchmark) matrices [2] that is modified by small deviations from benchmark and from exact QLC generated by the $\varepsilon$-parameter exponential factors.

## 4. Accurate neutrino mixing matrix as small shift from bimaximal benchmark mixing one

Benchmark neutrino mixing pattern describes bimaximal neutrino mixing at zero value of the parameter $\varepsilon = 0$ in terms of $\cos^2$-double-angle presentation [3]

$$\cos^2(2\theta_{sol}) = 0, \quad \cos^2(2\theta_{atm}) = 0, \quad \sin^2(2\theta_{13}) = 0, \qquad (23)$$

where $\theta_{sol} = \theta_{12}$ is the solar neutrino oscillation mixing angle, $\theta_{atm} = \theta_{23}$ is the atmospheric neutrino oscillation mixing angle and $\theta_{13}$ is the empirically small reactor neutrino oscillation mixing angle. From definition (23) follows the zero $\varepsilon$-approximation neutrino benchmark mixing matrix

$$\begin{pmatrix} 1/\sqrt{2} & 1/\sqrt{2} & 0 \\ -1/2 & 1/2 & 1/\sqrt{2} \\ 1/2 & -1/2 & 1/\sqrt{2} \end{pmatrix} \nu. \qquad (24)$$

It is the studied in literature [15] bimaximal neutrino mixing matrix.

At finite small parameter $\varepsilon$ the primary neutrino mixing conditions (23) for solar and atmospheric angles should be transformed to

$$\cos^2(2\theta_{sol}) = \varepsilon f'_1(\varepsilon), \quad \cos^2(2\theta_{atm}) = \varepsilon f'_2(\varepsilon), \qquad (25)$$

where $f'_i$, $i=1,2$, are finite functions of $\varepsilon$.

Close relations between quark and lepton flavor patterns is in the spirit of the SM. The accurate description of the world fit CKM mixing matrix by simple exponential correction factors to the zero $\varepsilon$-approximation matrix (3) is suggesting that similar description may exist for the lepton mixing matrix.

The condition of quark and neutrino equal deviations from the benchmark mixing matrices is equivalent to exact quark-lepton complementarity [5]. Thus, from comparison (31) with the above relations (21) and (22), final relations for realistic solar and atmospheric mixing angles follow:

$$\cos^2(2\theta_{sol}) = f(2\varepsilon)e^{-4\varepsilon} \cong 2\varepsilon \exp(-2\varepsilon), \quad \theta_{sol} \cong 34.04°, \qquad (26)$$

$$\cos^2(2\theta_{atm}) = f(\varepsilon^2)e^{-4\varepsilon} \cong \varepsilon^2 \exp(-4\varepsilon), \quad \theta_{atm} \cong 43°. \qquad (27)$$

The relations (26) and (27) with exponential factors included approximately obey the quadratic hierarchy flavor relation[4], $\cos^2(2\theta_{sol})/\cos(2\theta_{atm}) = 2\exp(-\varepsilon^2/2) \cong 2$. It is also an accurate relation between the empirical solar and atmospheric best-fit values (18) and (19) from oscillation data analysis [12].

At leading ε-approximation, the solar and atmospheric angles would be $\theta_{sol} \cong 33.05°$, and $\theta_{atm} \cong 42.65°$. The realistic deviations from leading ε-approximation are ~3% for solar and ~1% for atmospheric neutrino mixing angles.

If QLC were an exact regularity, the solar and atmospheric neutrino angles would be $\theta_{sol} \cong 31.95°$, and $\theta_{atm} \cong 42.64°$. The realistic deviations from exact QLC are ~6% for solar and ~1% for atmospheric neutrino mixing angles. Thus, in contrast to the common believe, the deviation of atmospheric neutrino mixing angle from exact QLC is about 6 times smaller than the solar one.

One may notice a more accurate geometric meaning of the universal deviation from exact QLC condition (20). At exact QLC, the neutrino and quark (1-2) and (2-3) mixing angles deviate equally from two orthogonal axes (respectively 'y-axis' and 'x-axis') in one quadrant in opposite directions:

---

[4] Quadratic hierarchy regularity for generic deviation-from-mass-degeneracy (DMD) flavor pairs is initially defined by the author in Phys. Lett. **B567**(2003)268 where it is used for the description of lepton mass ratios with the inference of quasi-degenerate neutrino type. It appears a general pattern in flavor mixing phenomenology and CP-violating phases at leading ε-approximation. The difference of the coefficients - '√2' for mass-ratios and '2' for mixing quantities – is less important and may be balanced by changing the powers of the relevant quantities.

$$\sin^2(\pi/2 - 2\theta_{sol}) = \sin^2(2\theta_c) = 2\varepsilon \exp 2\varepsilon,$$
$$\sin^2(\pi/2 - 2\theta_{atm}) = \sin^2(2\theta^q_{23}) = \varepsilon^2 \exp \varepsilon^2. \quad (28)$$

After deviation from exact QLC (20), the relative shifts of solar and atmospheric angles from exact QLC values are respectively given by

$$[(\pi/2 - 2\times 34.04°) - (\pi/2 - 2\times 31.95°)]/(\pi/2 - 2\times 31.95°) \cong -0.16,$$
$$[(\pi/2 - 2\times 43°) - (\pi/2 - 2\times 42.64°)]/(\pi/2 - 2\times 42.64°) \cong -0.15.$$

These values are close to the estimations in (20).

Realistic quark and neutrino mixing patterns differ not only by the main defining condition that the appropriate small deviations are from very different benchmark mixing patterns (3) and (24) at zero $\varepsilon$-approximation, but also by the universal QLC-violating condition (20) and quadratic hierarchy relations for (1-3) mixing angles and CP-violating phases beyond the leading $\varepsilon$-approximation.

The two other neutrino angles, 1) small reactor mixing angle $\theta_{13}$ and 2) Dirac CP-violating phase $\delta_\ell$, are determined by the considered quadratic DMD-hierarchy paradigm at leading $\varepsilon$-approximation for generic pairs of mixing quantities [3], namely

1) For the mixed quark-neutrino generic pair ($\theta^q_{13}$, $\theta^\ell_{13}$) –

$$\sin^2 2\theta^\ell_{13} = 2\sin 2\theta^q_{13}. \quad (29)$$

With $\sin 2\theta^q_{13}$ from (6) the result is

$$\sin^2 2\theta^\ell_{13} \cong 2\varepsilon^2 \cong 0.0135, \quad \theta^\ell_{13} \cong 3.3°. \quad (30)$$

**2)** For Dirac CP-violating phase $\delta_\ell$ –

$$\cos^2 \delta_\ell = 2\sin \delta_\ell, \quad \delta_\ell \cong (24.47°; 155.53°). \quad (31)$$

The two solutions $\delta_\ell \cong 24.47°$ and $\delta_\ell \cong 155.53°$ differ only by signs of $\cos \delta_\ell$. Notice, for both solutions $\sin \delta_\ell = (\sqrt{2}-1)$.

The inference $\sin^2 2\theta^\ell_{13} > 0.01$ means that lepton CP-violation may be discovered in coming neutrino oscillation experiments, e.g. [16].

Finally, the four angles (26), (27), (30) and (31) when used in the matrix form (7) predict an advanced form of the neutrino mixing matrix

$$V_{nu} \cong \begin{pmatrix} 0.827 & 0.559 & 0.058e^{-i\delta} \\ -0.409 - 0.033\,e^{i\delta} & 0.606 - 0.022e^{i\delta} & 0.681 \\ 0.382 - 0.035e^{i\delta} & -0.565 - 0.024e^{i\delta} & 0.730 \end{pmatrix} \quad (32)$$

with neutrino Dirac phase $\delta$ estimated in (31).

To summarize, the neutrino PMNS mixing matrix is inferred from the advanced quark CKM one by two leading regularities: i) universal deviation from exact QLC for the two largest mixing angles and ii) quadratic hierarchy equations for the Dirac CP-violating phases, and for the quark-neutrino mixed pair of two smallest mixing angles.

The Jarlskog CP-violating invariant [9] of the neutrino mixing matrix (32) for both values of Dirac CP-violating phases $\delta_\ell \cong 24.47°$ and $\delta_\ell \cong 155.53°$ is given by

$$J_\ell \cong 5.7 \times 10^{-3}, \quad (33)$$

- an observable quantity of lepton CP-violation.

## **6. A glimpse on new basic flavor physics**

The accurate quantitative results for particle mixing matrices suggest that emergence from empirical data of the new universal flavor constant $\varepsilon$ may imply a glimpse on new basic ingredient in flavor physics beyond the SM and the symmetry approach[5]. In this connection, it may be conceived that emergence of the $\varepsilon$-constant points to basic new flavor physics resembling the known emergence of the Plank constant h as guide to quantum mechanics. The Plank constant generates basic *local-nonlocal duality* phenomena of point-particle position states in the regular 3-dimensional space; it leads to a great many known simple and accurate low energy regularities in quantum mechanics. To detail the analogy, let us define the weak interaction eigenstates as local particle states in flavor space; than the particle mass eigenstates must be considered as nonlocal ones in this flavor space. By that definition, the ε-constant generates *local-nonlocal duality* phenomena of particle generation states in the discrete 3-flavor space. This duality may be the essence of basic flavor physics as the mentioned local-nonlocal point-particle duality is the essence of non-relativistic quantum mechanics. The new dimensionless ε-constant means a link between first particle generation and the two extra copy ones; it leads to the considered above simple and accurate low energy regularities in flavor mixing phenomenology. The limit $h \to 0$ would mean local classical mechanics with no quantum effects; the limit $\varepsilon \to 0$ would mean local-flavor first generation of elementary particles without observable extra copies in EW physics. Finally, like the h-constant in quantum mechanics, the ε-constant should be a crucial ingredient in flavor physics related not to flavor symmetry, but rather to flavor *dynamics* via connections to flavor *hierarchies.* As considered in previous arXiv-publications e.g. [3], the basic dimensionless particle flavor hierarchy quantities – *deviations of particle mass ratios and mixing angles from their benchmark values* – may likely be generated by the one universal ε-parameter and vanish in the limit $\varepsilon \to 0$.

---

[5] As known, the magnitude of the electric charge is beyond the gauge symmetry of the electromagnetic interactions; the magnitudes of the electroweak interaction constants are beyond the SU(2)xU(1) gauge symmetry, and beyond the mechanism of its spontaneous violation. In similar way, the $\varepsilon$-constant may be compatible with some flavor symmetry.

## 7. Conclusions

This paper is about exceptional quantitative effectiveness of the accurate empirical ε–parameter (1) in united quark-neutrino flavor mixing physics.

Below are listed some main results of the present research, which are interesting in the framework of particle flavor mixing phenomenology.

<1> The one-parameter quark mixing CKM model is based on relations of the three independent mixing angles in terms of one single exponential function of the ε-parameter. It appears in remarkably good agreement with the combined world SM-analyses of the quark CKM mixing matrix data.

It seems a glimpse on new basic flavor physics beyond the three flavor-copy Standard Model since for a physicist it would be hard to believe in an 'occasional' closely united meaningful quark-lepton flavor mixing system generated by one uniting empirically found ε-parameter; that system is in highly accurate quantitative agreement with the experimental data[6] and its structure is simple and transparent. It would be too apprehensive to easily 'explain' that system as a mystical conglomerate of coincidences[7]. At physics frontier, 'a glimpse on new basic flavor physics' is a realistic and much simpler idea.

---

[6] Plus another fitting system of lepton and quark mass ratios in terms of that ε-parameter [3].
[7] One critic without denying the results uses the term "numerology" instead, but evidently it is not a less mystical 'explanation'.

Both the presentation (8) for the quark mixing matrix and the inference for neutrino mixing matrix (32) are factual results reproducing those from data analyses - at least.

The CKM and PMNS mixing matrices are fully represented in analytical form by equations (6), (9) for quarks and (26), (27), (29), (30) for neutrinos respectively.

<**2**> The status of the QLC idea [5] is actually advanced by the new observation of *equal* solar and atmospheric deviations from exact QLC at next to leading $\varepsilon$-approximation. And the more so, as the magnitude of that deviation is quantitatively estimated via the $\varepsilon$-parameter and used for accurate calculations of neutrino mixing quantities. Notice that the observation of universal deviation from exact QLC was made possible by the analytical form of the CKM mixing matrix in equations (6) and (9). This observation enhanced the interesting QLC hint (see e.g. recent review [17]) to the level of predictive accurate quantitative QLC-phenomenology supported by known experimental data.

The initial defining form of QLC is at zero $\varepsilon$-approximation, benchmark matrices. At leading nonzero $\varepsilon$-approximation (i.e. without exponential corrections) exact QLC is maintained by equal deviations from benchmark mixing of quarks and neutrinos for both pairs of quark and neutrino (1-2) and (2-3) mixing parameters (comp. the first two relations in (6) for quarks with (26), (27) for neutrinos).

The QLC condition is evidently not related to the small (1-3) mixing angles. This is an inference from zero (1-3) angles in both benchmark matrices (3) and (24), unlike the other angles, and smallness of the ε-parameter. Probably, the two small mixing angles $θ^q_{13}$ and $θ^{nu}_{13}$ make up one mixed quark-neutrino generic pair and therefore obey the quadratic hierarchy paradigm.

All elements of the neutrino mixing matrix are obtained from quark mixing ones by two observed quark-lepton regularities: i) universal deviation from exact explicit QLC for the two largest (1-2) and (2-3) mixing angles, and ii) quadratic hierarchy flavor rule for Dirac CP-violating phases [4a], and the neutrino and quark (1-3) mixing angles.

Both QLC and quadratic hierarchy are exact flavor regularities at zero and leading nonzero ε-approximations. At next to leading ε-approximation they are weakly violated by exponential ε-factors, which lead to the discussed accurate results in flavor quark-neutrino mixing phenomenology.

<**3**> The Kobayashi-Maskawa CP-violating phase [6] is an empirical parameter in the CKM matrix, but otherwise $δ$ is arbitrary and $δ=0$ is not forbidden by the phenomenology. An interesting problem of the CKM mixing matrix is to find a definite phenomenological rule for choosing one specific value of the CP-violating phase $δ$.

Since the real and imaginary parts of the CP-violating terms $e^{iδ}$, which is possible in the three generation particle mixing matrices, make up 'generic' flavor pairs, they should obey the quadratic hierarchy relation. It provides the missing physical condition of necessary CP-violation in the quark and neutrino (Dirac) mixing matrices. Thus, it is again the flavor quadratic hierarchy paradigm that makes CP-violation in the CKM (and PMNS) matrix compulsory and predicts finite CP-violating phase $\delta_q$ (and $\delta_\ell$) in preliminary agreement with basic facts in heavy meson physics.

The derived value of quark CP-violating phase is compatible with heavy flavor experimental data analyses of the unitarity angles [7]. The estimated angles may be helpful for determining quark CP-violating phase $\delta_q$ from coming accurate experimental data on unitarity triangle angles from B-factories and LHCb experiments. Especially, the inferred accurate closeness, to within ~$6 \times 10^{-4}$, of the magnitudes of the unitarity triangle gamma-angle $\gamma \cong 65.49°$ and quark CP-violating phase $\delta_q \cong 65.53°$ is emphasized. If confirmed by coming experimental data, it points to an interesting regularity that provides necessary CP-violation in flavor mixing physics.

It should be noticed that if the prediction for quark CP-violating phase $\delta_q \cong 65.53°$ will be confirmed by new data, the related value of the neutrino Dirac CP-violating phase $\delta_\ell \cong 24.47°$ (or $155.53°$) will be also partly supported since both phases are solutions of the same quadratic hierarchy equation[8] with complementarity relations $24.47° = \pi/2 - \delta_q$ and $155.53° = \pi/2 + \delta_q$.

---

[8] Notice that all considered above dimensionless flavor quantities are grouped in pairs that obey either the quadratic hierarchy condition or the quark-lepton complementarity one. But there is one exception – the real and imaginary parts of the two CP-violating phases {$\cos\delta_q$, $\sin\delta_q$}

**References**

[1] S. L. Glashow, arXiv:0912.4976.

[2] W. Rodejohann, Phys. Rev. **D69**, 033005 (2004); R. N. Mohapatra, A. Yu. Smirnov, Ann. Rev. Nucl. Part. Sci., **56**, 569 (2006).

[3] E. M. Lipmanov, arXiv:0812.1550; arXiv:0902.1486; arXiv:0906.1982.

[4] E. M. Lipmanov, arXiv:0909.4257.

[4a] E.M. Lipmanov, arXiv:0710.4125.

[5] A. Yu. Smirnov, hep-ph/0402264; M. Raidal, Phys. Rev. Lett., **93**, 161801 (2004); H. Minakata, A. Yu. Smirnov, Phys. Rev.,**D70**, 073009 (2004); P. H. Frampton, R.N. Mohapatra, hep-ph/0407139; A.Datta, L. Everet, P. Ramond, hep-ph/0503222; Z. z. Xing, hep-ph/0503200; J. Ferrandes, S. Pakvasa, Phys. Rev. **D71**, 033004 (2005); S. Antush, S. F. King, R. N. Mohapatra, Phys. Lett., **B618**, 150 (2005); also S. T. Petcov, A. Yu. Smirnov, Phys. Lett. **B322**, 109 (1994); and especially W. Rodejohann, hep-ph/0309249 and [2].

[6] N. Cabibbo, Phys. Rev. Lett., **10** (1963) 531; M. Kobayashi, K. Maskawa, Prog. Theor. Phys. **49** (1973) 652.

[7] Particle Data, C.Amsler et al. Phys.Lett.**B667**,1(2008).

[8] S. Luo, Z-z. Xing, arXiv:0912.4593.

[9] C. Jarlskog, Phys. Rev. Lett., **55**, 1039 (1985).

[10] SNO Collab., B. Aharmin et al, arXiv:0910.2984

[11] KamLAND Collab., S. Abe et al, arXiv:0801.4589.

[12] G. L. Fogli et al., Nucl.Phys.Proc.Suppl.188:27,2009.

[13] P. F. Harrison, D. H. Perkins, W.G. Scott, Phys.

---

and $\{\cos\delta_\ell, \sin\delta_\ell\}$ are complementary and generic pairs at the same time. This important double condition makes the CP-violation in the quark and neutrino mixing matrices compulsory.

Lett. B530 (2002) 167.

[14] M. C. Gonzalez-Garcia, M. Maltoni, J. Salvado, arXiv:1001.4524; T. Schwetz, M.A. Tortola, J.W.F. Valle, New J. Phys. 10:113011,2008.

[15] F. Vissani, hep-ph/9708483; V. D. Barger, S. Pakvasa, T. J. Weiler, K.Wisnand, Phys.Lett., **B437**, 107 (1998); H. Georgi, S.L. Glashow, Phys.Rev.,**D61**, 097301 (2000).

[16] A. Rubbia, arXiv:0908.1286.

[17] G. Altarelli and F. Feruglio, arXiv:1002.0211.